\documentclass[showpacs,aps,amsmath,amssymb]{revtex4}

\usepackage{CJK}

\usepackage[latin1]{inputenc}
\usepackage{graphicx}
\usepackage{bm}
\usepackage{float}
\usepackage{xcolor}
\usepackage{array}
\usepackage{braket}
\usepackage{mathtools}
\usepackage{microtype}
\usepackage{wasysym}
\renewcommand{\baselinestretch}{1.0}
\begin{document}

\renewcommand{\baselinestretch}{1.0}
\title{Shortcuts to adiabatic population inversion via time-rescaling: stability and thermodynamic cost}

\author{Jair da Silva Andrade, \^{A}ngelo F. da Silva Fran\c{c}a, and Bert\'{u}lio de Lima Bernardo}

\affiliation{Departamento de F\'{\i}sica, Universidade Federal da Para\'{\i}ba, 58051-900 Jo\~ao Pessoa, PB, Brazil}

\begin{abstract}

A shortcut to adiabaticity (STA) is concerned with the fast and robust manipulation of the dynamics of a quantum system which reproduces the effect of an adiabatic process. In this work, we use the time-rescaling method to study the problem of speeding up the population inversion of a two-level quantum system, and the fidelity of the fast dynamics versus systematic errors in the control parameters. This approach enables the generation of shortcuts from a prescribed slow dynamics by simply rescaling the time variable of the quantum evolution operator. It requires no knowledge of the eigenvalues and eigenstates of the Hamiltonian and, in principle, no additional coupling fields. From a quantum thermodynamic viewpoint, we also demonstrate that the main properties of the distribution of work required to drive the system along the shortcuts are unchanged with respect to the reference (slow) protocol.                                                    
   
\end{abstract}

\maketitle


The precise control of the dynamics of two-level quantum systems with time-dependent interacting fields is essential to the manipulation of the internal degrees of freedom of atoms and molecules, having applications in many emerging quantum technologies \cite{wiseman,vand,torre,brif,bernardo1,frank,zhou,chu}. In this case, when the task is to attain a complete population inversion, the most common options are the usage of a resonant $\pi$ pulse, composite pulses, and adiabatic passage protocols \cite{allen,vit}. The $\pi$ pulse and composite pulse methods can provide fast inversions if high-intensity fields are used, but these techniques are very sensitive to imperfections in the intensity and the area of the pulse \cite{allen}. In this respect, adiabatic processes become an interesting alternative for being robust to experimental imperfections, but at the same time they are vulnerable to decoherence due to the long evolution times required \cite{guery}. In order to remedy the drawbacks of these approaches, the strategy is to devise quantum dynamics that are both fast and robust. Over the last decade, several techniques have been developed that allowed the robust manipulation of quantum systems at timescales shorter than decoherence times; the so-called ``shortcuts to adiabaticity'' (STA) \cite{guery,torrengui}.

The STA methods aim to obtain analytic solutions describing fast and robust quantum time evolutions capable of reproducing the same effect of adiabatic transformations. They have been used to speed up the dynamics of harmonic potentials \cite{xchen,accon,bernardo}, spin systems \cite{berry,bernardo}, as well as two- and three-level atomic systems \cite{chen,chen2,kiely,song1,song2}. These methods have also been experimentally used to control the dynamics of Bose-Einstein condensates in optical lattices \cite{bason}, trapped ions \cite{an}, nitrogen-vacancy centers \cite{zhou}, and superconducting qubits \cite{wang}. Among the most popular and well-studied STA methods, we can mention the inverse engineering approaches using the Lewis-Riesenfeld (LR) invariants \cite{chen3}, the counterdiabatic driving (CD) \cite{berry,demi,demi2,deffner}, and the fast-forward (FF) technique \cite{masuda1,masuda2,torrengui2}. Recently, a new STA approach was proposed by one of us, called time-rescaling method \cite{bernardo}, which consists of rescaling the time variable in the quantum evolution operator in order to find fast dynamics that enable    reproducing the same final state and external conditions of a prescribed (reference) quantum process, be it adiabatic or not. This method is independent of the eigenspectrum and eigenstates of the Hamiltonian, and has also been applied to study the control of a Dirac dynamics \cite{roy}.  We can also mention other STA schemes in which knowledge about the eigenvalues and eigenstates of the adiabatic Hamiltonian is not required \cite{sels,hchen,ran}

In this work, we report on the application of the time-rescaling method to realize fast population inversion processes in a two-level quantum system, taking the  Allen-Eberly (AE) adiabatic
passage scheme as the reference dynamics \cite{allen,chen,vitanov}.
We show that the robustness of the method is equivalent to other STA techniques, and that the work distribution properties of the fast dynamics are unchanged with respect to the reference (slow) process. Some key benefits can be pointed out. First, the method to engineer the Hamiltonian that generates the fast dynamics is totally analytic. Second, contrary to the usual CD method \cite{chen,rusch}, the time-rescaling requires only one laser to implement the shortcuts. In the following section we review the time-rescaling method. In Sec.~III we address the problem of population transfer in a two-level system and show the behavior of the control parameters and the populations as a function of time. Sec.~IV discusses the stability of the dynamics upon systematic errors in the control parameters. In Sec.~V we then demonstrate a general result on
the distribution of work
required to drive the system
along the shortcuts obtained via time-rescaling. The conclusion is presented in Sec.~VI.    

\section*{METHODS}

\subsection*{Time-Rescaling Method}

In this section we review the time-rescaling method of quantum control \cite{bernardo}. The dynamics of a closed quantum system subject to a time dependent Hamiltonian $\hat{H}(t)$, acting between an initial time $0$ and a final time $t_{f}$, is described by an unitary time evolution operator $\hat{U}(t)$, which satisfies the Schr\"{o}dinger equation,
\begin{equation}
\label{1}
\hat{H}(t)\hat{U}(t)=i \hbar \frac{\partial}{\partial t} \hat{U}(t).
\end{equation}
The general solution to this equation, with the initial condition $\hat{U}(0) = \hat{\mathcal{I}}$, where $\hat{\mathcal{I}}$ is the identity operator, is given by \cite{sakurai} 
\begin{equation}
\label{2}
\hat{U}(t_{f}) = \hat{\mathcal{T}}\exp \left\{ -\frac{i}{\hbar} \int_{0}^{t_{f}} \hat{H}(t) dt \right\},
\end{equation}
where $\hat{\mathcal{T}}$ is the time-ordering operator. For convenience, let us call this time evolution generated by $\hat{H}(t)$ the {\it reference} dynamics. 

Now suppose we want to devise a new dynamics that, starting from the same initial state, reproduces the same final state of the system, but in a shorter time interval ($\Delta t < t_{f}$). Needless to say that the generator of this new dynamics cannot be $\hat{H}(t)$. In order to accomplish this task, we make a simple change in the time variable in Eq.~(\ref{2}) according to the relation $t=f(\tau)$, so that we obtain the following time evolution operator
\begin{align}
\label{3}
\hat{\mathcal{U}}(\Delta t) &= \hat{\mathcal{T}} \exp \left\{ -\frac{i}{\hbar} \int_{f^{-1}(0)}^{f^{-1}(t_{f})} \hat{H}[f(\tau)]f'(\tau) d\tau \right\} \nonumber \\
&= \hat{\mathcal{T}} \exp \left\{ -\frac{i}{\hbar} \int_{f^{-1}(0)}^{f^{-1}(t_{f})} \hat{\mathcal{H}}(\tau) d\tau \right\},
\end{align}
where $\Delta t = f^{-1}(t_{f}) - f^{-1}(0)$. The operator
\begin{align}
\label{3.1}
\hat{\mathcal{H}}(\tau) = \hat{H}[f(\tau)]f'(\tau)
\end{align}
is the Hamiltonian of the new time evolution that is called {\it time-rescaled} (TR) dynamics. The functions $f'(\tau)$ and $f^{-1}(\tau)$ are the first derivative and the inverse of $f(\tau)$, respectively. Since Eq.~(\ref{3}) was obtained from Eq.~(\ref{2}) by a simple change of variable, it turns out that the resulting action of the TR evolution operator after a time $\Delta t$ is equal to that of the reference evolution operator after a time $t_{f}$, i.e., $\hat{\mathcal{U}}(\Delta t) = \hat{U}(t_{f}) $. This means that if these operators are applied to an arbitrary initial state $\ket{\psi(0)}$ of the system, they will produce precisely the same final state, $\ket{\psi(t_{f})}= \hat{U}(t_{f}) \ket{\psi(0)} = \hat{\mathcal{U}}(\Delta t) \ket{\psi(0)}$. 

From the application point of view, besides the correspondence between the initial and final states of the system, it is also important that the reference and TR Hamiltonians are the same at the beginning and the end of the respective processes. This guarantees that, in case the system is in a stationary state at the beginning and the end of the reference dynamics, the same holds true for the TR protocol. This property is central for quantum control purposes \cite{wiseman,brif}. The crucial point to conceive a TR dynamics that fulfills these requirements, while being also faster than the reference one, lies in choosing an appropriate time-rescaling function $f(\tau)$. In short, this function must be such that the following four properties are satisfied: (i) the initial times of the reference and TR dynamics are the same: $f^{-1}(0) = 0$, (ii) the TR dynamics must be faster:  $f^{-1}(t_{f})<t_{f}$, (iii) the initial Hamiltonians must be the same:  $\hat{\mathcal{H}}[f^{-1}(0)] = \hat{H}(0)$, and (iv) the final Hamiltonians must be the same: $\hat{\mathcal{H}}[f^{-1}(t_{f})]=\hat{H}(t_{f})$. As can be seen, it is trivial to find a function that fulfills the requirements (i) and (ii). By contrast, (iii) and (iv) are fulfilled given that $f'[f^{-1}(0)] = f'[f^{-1}(t_{f})] = 1$.

It was shown in Ref. \cite{bernardo} that a
simple and pragmatic function that fulfills all the above properties is  
\begin{equation}
\label{4}
f(\tau) = a \tau - \frac{(a-1)}{2 \pi a} t_{f} \sin \left( \frac{2 \pi a}{t_{f}} \tau\right).
\end{equation}
In this case, the inverse function $f^{-1}(\tau)$ cannot be expressed exactly in terms of standard functions, but the properties $f^{-1}(0) = 0$, $f^{-1}(t_{f}) = t_{f}/a$, $f'(0) = 1$ and $f'(t_{f}/a) = 1$ hold exactly, as can be directly checked. With this, we have that $f(\tau)$ given in Eq.~(\ref{4}) is an appropriate time-rescaling function. That is to say that when this function is used in Eq.~(\ref{3}) with $a>1$, the TR dynamics transforms any initial state of the system into the corresponding final state that results from the reference dynamics. In the end, the TR protocol sets the system with same final state and external conditions as the reference one, but in a time interval which is $a$ times shorter. For this reason $a$ is called {\it time contraction parameter}. Before closing this section, we call attention to two important points. First, we have that if the reference dynamics is adiabatic, the TR dynamics works as a STA. Second, if we choose $0<a<1$, the TR dynamics takes more time than the reference one to produce the final state and external conditions. In principle, this $0<a<1$ case has no apparent technological application.

\section*{RESULTS}

\subsection*{Population Inversion in a Two-Level System}

The time-rescaling scenario was used in Ref.~\cite{bernardo} to investigate how to accelerate the dynamics of the parametric oscillator, the transport of a particle in a harmonic trap, and the spin-1/2 system in a magnetic field. Here, we use the method to address the problem of speeding up the adiabatic passage in a two-level atomic system. Specifically, we will be interested in creating a shortcut to the adiabatic inversion of the populations of two atomic levels denoted by $\ket{1}$ and $\ket{2}$. This problem has been addressed using the transitionless
CD algorithm and the inverse engineering approach \cite{chen,chen2}. In this case, the two methods were shown to provide similar nonadiabatic
shortcuts. 

The time-dependent Hamiltonian that generates the adiabatic passage in a two-level atomic system interacting with a laser can be written in the basis $\{\ket{1},\ket{2}\}$ as 
\begin{align}
\hat{H}(t) = \dfrac{\hbar}{2} \begin{pmatrix}
\Delta & \Omega_{R} e^{i \varphi} \\ 
\Omega_{R} e^{-i \varphi} & -\Delta
\end{pmatrix},
\label{eq:1}
\end{align}
where we have the time-dependent functions $\Omega_{R} = \Omega_{R}(t)$ representing the Rabi frequency, $\Delta = \Delta (t)$ the detuning, and $\varphi(t)$ a
time-dependent phase. While $\Omega_{R}$ depends on the atomic transition dipole moment and the electric field of the laser pulse, $\Delta$ is the difference between the atom transition frequency $\omega_{0}$ and the laser frequency $\omega_{L}$. The instantaneous eigenkets of the Hamiltonian are given by
\begin{align}
|n_{+}(t)\rangle &= \sin \left(\dfrac{\theta}{2}\right)|1\rangle + e^{i \varphi} \cos \left(\dfrac{\theta}{2}\right) |2\rangle, \\
|n_{-}(t)\rangle &= - e^{- i \varphi} \cos \left(\dfrac{\theta}{2}\right)|1\rangle + \sin \left(\dfrac{\theta}{2}\right) |2\rangle,
\label{eq:2}
\end{align}
with $\theta = \theta (t) \equiv \text{arccos}(\Delta/\Omega)$. Here, the eigenenergies are  $E_{\pm}(t) = \pm \hbar \Omega (t)/2$, with $\Omega = \sqrt{\Delta^{2}+\Omega^{2}_{R}}$ being the so-called generalized Rabi frequency. If we assume the initial state of the system as $|\psi_{\pm}(0)\rangle = |n_{\pm}(0)\rangle$,
the condition of adiabaticity \cite{chen,chen2}, $\left|\Omega_{R}\dot{\Delta} - \dot{\Omega}_{R}\Delta \right| \ll \left|\Omega^{3} \right|$, and $\varphi = 0$, the time evolution will be described
according to the adiabatic theorem \cite{sakurai}
\begin{align}
\ket{\psi_{\pm}(t_{f})} = \text{exp}\left\{-\dfrac{i}{\hbar}\int_{0}^{t_{f}} E_{\pm}(t) dt  \right\} \ket{n_{\pm}(t)},
\end{align}
which guarantees the absence of transitions.

By choosing the initial state as $|n_{+}(0)\rangle$, the adiabatic time evolution of the populations of the levels 1 and 2 are found to be
\begin{equation}
\label{pad1}
P_{1}^{\text{ad}}(t) = |\langle 1|n_{+}(t)\rangle|^{2} = \sin^{2}\left(\dfrac{\theta}{2}\right),
\end{equation}
\begin{equation}
\label{pad2}
P_{2}^{\text{ad}}(t) = |\langle 2|n_{+}(t)\rangle|^{2} = \cos^{2}\left(\dfrac{\theta}{2}\right).
\end{equation}
Here, we shall consider the AE adiabatic passage scheme, whose manipulation of the Rabi frequency and detuning can be prescribed, respectively, as follows \cite{chen,allen,vitanov}:
\begin{equation}
\label{omegaR}
\Omega_{R}(t) = \Omega_{0}\ \text{sech} \left[\dfrac{\pi (t-4t_{0})}{2t_{0}} \right],
\end{equation}
\begin{equation}
\label{delta}
\Delta(t) = \left(\dfrac{2\beta^{2}t_{0}}{\pi} \right)\tanh \left[\dfrac{\pi (t-4t_{0})}{2t_{0}} \right],
\end{equation}
where the parameters $\Omega_{0}$ and $\beta$ are real constants with dimension of frequency, and $t_{0}$ is a constant representing the characteristic time scale of the dynamics. We set the above equations in a form that the protocol initiates at a time $t_{i} = 0$ and ends at $t_{f} = 8 t_{0}$. In theory, the AE scheme requires an infinitely long time to provide a population inversion with unit probability. However, using the scheme with a total time duration of $8 t_{0}$, as described in Eqs.~(\ref{omegaR}) and~(\ref{delta}), a successful population inversion is achieved with probability higher than 0.999.

The time behavior of $\Omega_{R}(t)$ and $\Delta(t)$ are shown in Fig.~1(a), where we chose  $\Omega_{0} = 2$, $\beta = \sqrt{2}$ and $t_{0} = 1$. At $t = 0$ we have that $\Omega_{R} \approx 0$ and $\Delta \approx - 4/ \pi$, which provides $\theta(0) \approx \pi$. With this, the chosen initial state $\ket{n_{+}(0)}$  becomes approximately the lower energy eigenstate $\ket{1}$. Accordingly, the corresponding time evolutions of the populations calculated with basis on Eqs.~(\ref{pad1}) and (\ref{pad2}) are shown in Fig.~1(b). 
\\

\begin{figure}[H]
\centering
\includegraphics[scale=0.28]{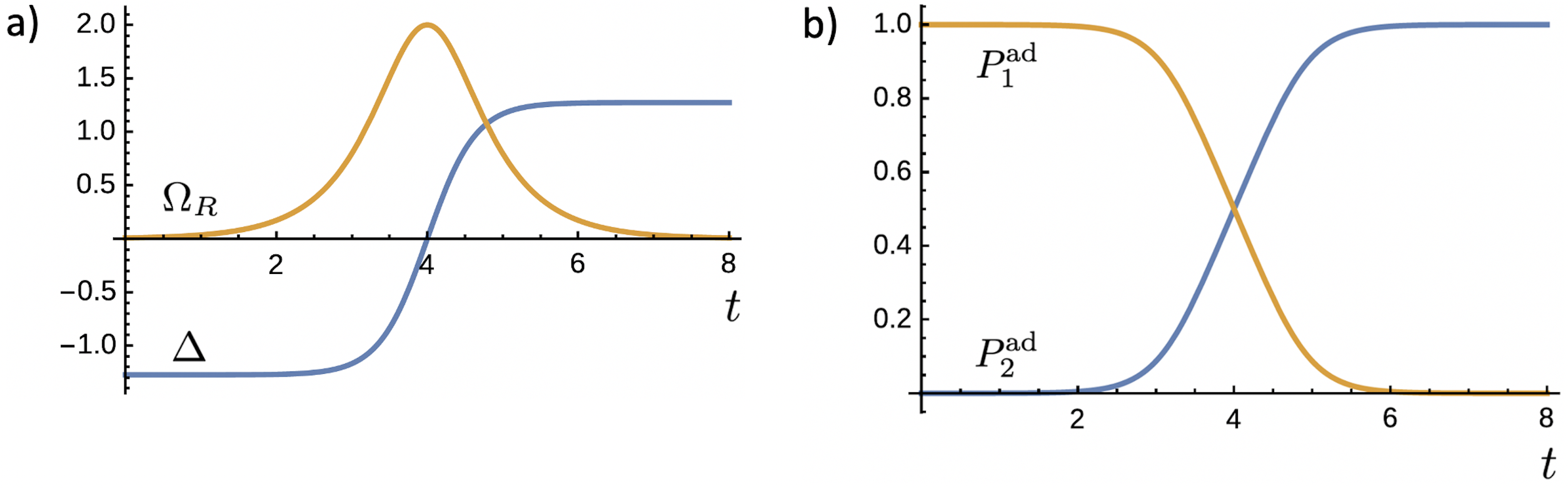}
\caption{Time behavior of: a) $\Omega_{R}$ and $\Delta$ for the AE adiabatic passage scheme as described in Eqs.~(\ref{omegaR}) and (\ref{delta}), and b) the populations of the energy levels with the initial state given by $\ket{n_{+}(0)} \approx \ket{1}$. Parameters: $\Omega_{0} = 2$, $\beta = \sqrt{2}$, and $t_{0} = 1$. }
\label{fig:f1}
\end{figure}

We now turn our attention to the problem of accelerating the inversion of population process in the two level system by using the time-rescaling method. In doing so, we consider the time evolution generated by the Hamiltonian of Eq.~(\ref{eq:1}) with $\varphi = 0$ and the control parameters varying according to Eqs.~(\ref{omegaR}) and (\ref{delta}) as the reference dynamics. As such, the TR Hamiltonian that generates the fast process is given by $\hat{\mathcal{H}}(\tau) = \hat{H}[f(\tau)]f'(\tau)$, shown in Eq. (\ref{3.1}), with $f(\tau)$ given as in Eq.~(\ref{4}). This fast Hamiltonian is then found to be
\begin{equation}
\hat{\mathcal{H}} (\tau)= \dfrac{\hbar}{2} \begin{pmatrix}
\tilde{\Delta} & \tilde{\Omega}_{R}\\ \tilde{\Omega}_{R} & -\tilde{\Delta}
\end{pmatrix},
\label{eq:8}
\end{equation}
with $\tilde{\Delta} = f'(\tau) \Delta[f(\tau)]$ and $\tilde{\Omega}_{R} = f'(\tau) \Omega_{R}[f(\tau)]$. In this form, the TR Rabi frequency and detuning are given explicitly by
\begin{widetext}
\begin{equation}
\label{rabitilde}
\tilde{\Omega}_{R} (\tau) = \Omega_{0} \left[  a - (a-1) \cos \left( \dfrac{2\pi a}{t_{f}} \tau \right) \right] \text{sech} \left\{\dfrac{\pi}{2t_{0}} \left[a \tau - \dfrac{(a-1)}{2\pi a} t_{f} \sin \left(\dfrac{2\pi a}{t_{f}} \tau \right) -4t_{0} \right] \right\},
\end{equation}
\begin{equation}
\label{detuningtilde}
\tilde{\Delta}(\tau)  = \left(\dfrac{2\beta^{2}t_{0}}{\pi} \right) \left[  a - (a-1) \cos \left( \dfrac{2\pi a}{t_{f}}\tau \right) \right] \tanh \left\{\dfrac{\pi}{2t_{0}}\left[ a\tau - \dfrac{(a-1)}{2\pi a} t_{f} \sin\left(\dfrac{2\pi a}{t_{f}}\tau \right) -4t_{0} \right] \right\}.
\end{equation}
\end{widetext}
In Fig.~2 we display the time behavior of $\tilde{\Omega}_{R}$ and $\tilde{\Delta}$ for $a = 2$ and $a = 10$. The peak of the Rabi frequency $\tilde{\Omega}^{max}_{R}$ always takes place at $\tau = t_{f}/2a$, which if substituted into Eq.~(\ref{rabitilde}) yields $\tilde{\Omega}^{max}_{R} = (2a-1) \Omega_{0} $. This means that a TR dynamics which is $a$ times faster than the adiabatic process requires a Rabi frequency peak $2a-1$ times greater.

\begin{figure}[h!]
\centering
\includegraphics[scale=0.30]{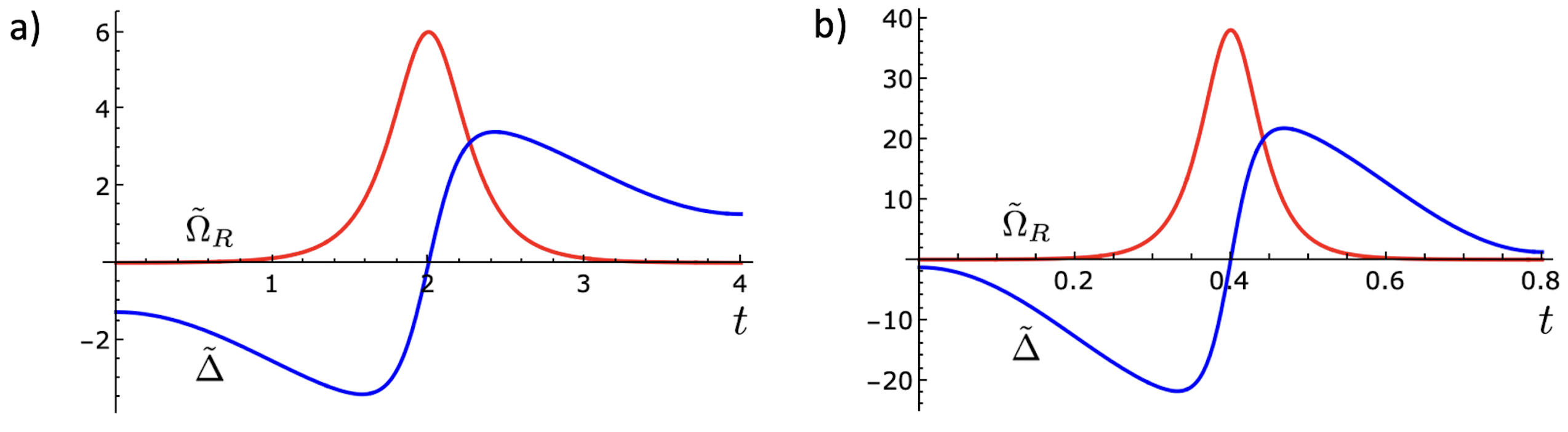}
\caption{Behavior of the TR functions $\tilde{\Omega}_{R}$ and $\tilde{\Delta}$ with the contraction parameter assuming the values a) $a=2$ and b) $a=10$. In both cases we note that the dynamics occur $a$ times faster than the reference control process described in Fig. 1(a), with the same initial and final conditions. Same parameters as in Fig. 1.}
\label{fig:f1}
\end{figure}

To investigate whether the fast dynamics generated by the Hamiltonian of Eq.~(\ref{eq:8}), with the control parameters determined by Eqs.~(\ref{rabitilde}) and~(\ref{detuningtilde}), is effective to accomplish the population inversion, we cannot use the results obtained for the adiabatic time evolution of the populations, as described in Eqs.~(\ref{pad1}) and~(\ref{pad2}). This is because the fast dynamics are not expected to satisfy the adiabatic approximation. Therefore, the time evolution of the populations must be calculated by directly solving the Schr\"{o}dinger equation with the TR Hamiltonian,
\begin{equation}
\label{seq}
\hat{\mathcal{H}}(\tau) \ket{\psi_{1,2}(\tau)}= i \hbar \frac{\partial}{\partial \tau} \ket{\psi_{1,2}(\tau)}, 
\end{equation}
with the initial condition $\ket{\psi_{1}(0)} = \ket{1}$. In this case, the evolution of the populations of the levels 1 and 2 are given by:
\begin{equation}
\label{ptr1}
P_{1}^{\text{tr}}(\tau) = |\langle 1|\psi_{1}(\tau) \rangle |^{2},
\end{equation}
\begin{equation}
\label{ptr2}
P_{2}^{\text{tr}}(\tau) = |\langle 2|\psi_{1}(\tau) \rangle |^{2}.
\end{equation}
We numerically solved Eq.~(\ref{seq}) and calculated the evolution of the populations of the levels. In Fig. 3 we show the results for $a=2$ and $a=10$. 

As can be seen, when the dynamics of the system is driven according to the TR protocol described by Eqs. (\ref{rabitilde}) and (\ref{detuningtilde}), the population inversion is successfully realized $a$ times faster than in the reference ($a = 1$) case of Eqs. (\ref{omegaR}) and (\ref{delta}). This certifies the suitability of the time-rescaling method in providing shortcuts to adiabatic passage for  two-level quantum systems. Still, some important remarks must be made. First, our calculations showed that the qualitative profile of the population inversions observed in Fig. 3 is independent of the contraction parameter $a$. Thus, similar to the CD and inverse engineering methods, the time-rescaling protocol also works for arbitrarily short times. Second, from the theoretical side, the time-rescaling approach has the advantage that the time-dependence of the control parameters can be expressed analytically as shown in Eqs.~(\ref{rabitilde}) and~(\ref{detuningtilde}); in general this is not the case of the other methods \cite{guery}. Third, the experimental manipulation of the control parameters as described here can be implemented using only one laser with a varying detuning. This is an important advantage of the time-rescaling scheme over the CD, which in general requires two lasers with the same frequency, orthogonal polarization, and time-dependent intensities, but different intensity shapes \cite{chen,rusch}.

\begin{figure}[h!]
\centering
\includegraphics[scale=0.30]{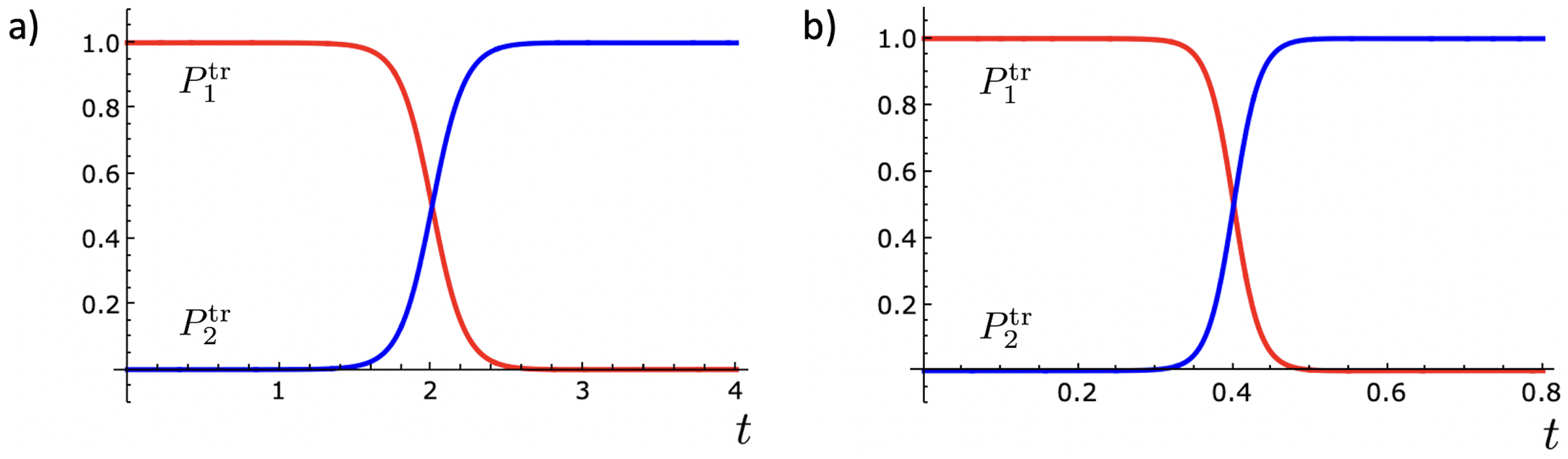}
\caption{Time evolution of the populations for the TR passage protocol with the initial state given by $\ket{1}$, and the contraction parameter assuming the values a) $a=2$ and b) $a=10$. In both cases we observe that the population inversion occurs at least $a$ times faster than in the reference ($a = 1$) protocol, shown in Fig. 1(b). Same parameters as in Fig. 1.}
\label{fig:f1}
\end{figure}

\subsection*{Stability against systematic errors}

In practical applications, STA protocols must be fast to avoid the influence of environment-induced decoherence on the system, but it is also essential that the fast dynamics be robust and stable to unavoidable errors in the manipulation of the control parameters. Let us now examine the stability of the fast population inversion method described in the previous section when systematic errors in the Rabi frequency and the detuning are considered. Fig. 4 displays the behavior of the fidelity $F = P_{2}$, which is the probability of successfully obtaining the state $\ket{2}$ from the initial state $\ket{1}$ after realizing the population inversion protocol, with respect to changes in the Rabi frequency and detuning. It shows how $F$ varies when: a) $\Omega_{0}$ changes to $\Omega_{0}(1+\epsilon)$, and b) $\beta^2$ changes to $\beta^2(1+\delta)$. In both cases the maximum fidelity ($F=1$) do not correspond exactly to the points $\epsilon = 0$ and $\delta = 0$. This is because the reference protocol, and hence the derived fast protocols, attain the population inversion with success probability higher than 0.999 (but not 1), as already discussed in the previous section. All results were found to be independent of the contraction parameter $a$. That is, the fidelity in the time-rescaling method is not modified with respect to errors in the parameters $\Omega_{0}$ and $\beta^2$ when the time duration of the protocol is shortened. 

\begin{figure}[h!]
\centering
\includegraphics[scale=0.30]{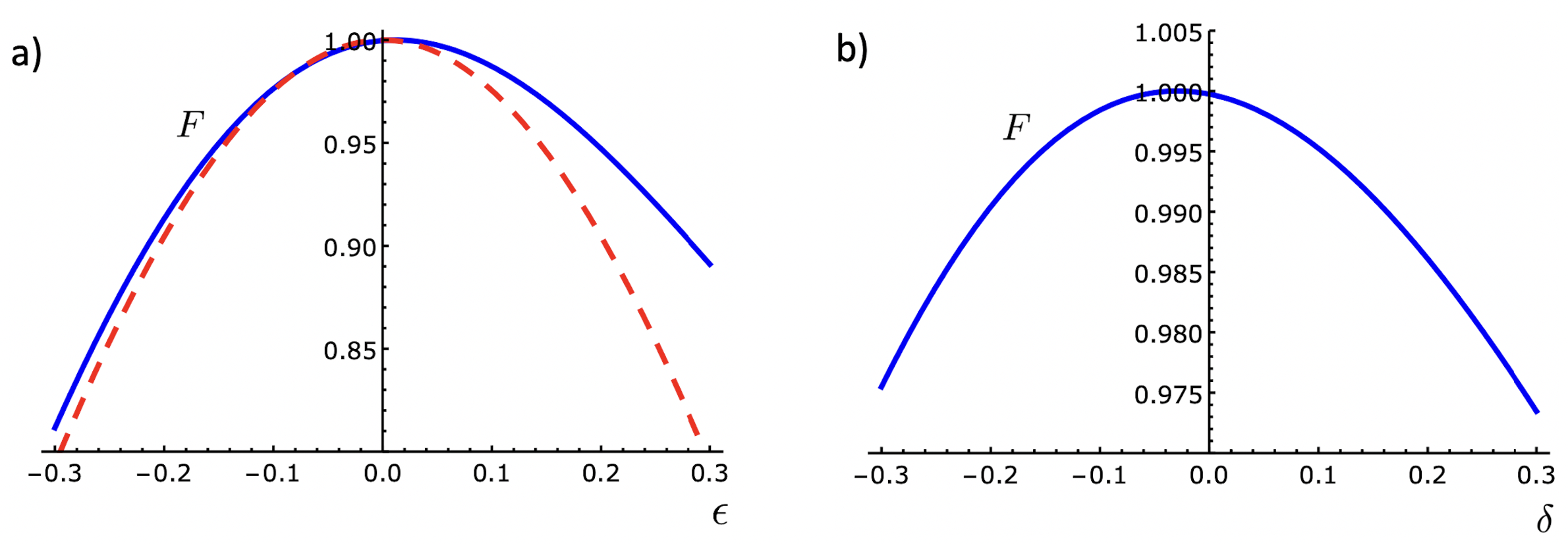}
\caption{Behavior of the fidelity ($F=P_{2}$) versus systematic errors in: a) the Rabi frequency, $\Omega_{0} \rightarrow \Omega_{0}(1+\epsilon)$, and b) the detuning, $\beta^2 \rightarrow \beta^2(1+\delta)$, for the TR dynamics (solid blue lines). The results are independent of the contraction parameter $a$. For comparison, in a) we show the behavior of $F$ in the square $\pi$ pulse case versus a systematic error in the Rabi frequency, $\Omega_{R} \rightarrow \Omega_{R}(1+\epsilon)$ (dashed red line). Same parameters as in Fig. 1.}
\label{fig:f1}
\end{figure}

In general, the protocols discussed here are more susceptible to errors in the Rabi frequency than the detuning because atoms in different positions experience different fields, due to, for example, a Gaussian shape laser. In Fig.~4(a) we can observe that errors of up to 20\% in $\Omega_{0}$ lead to $F \geq 0.913$ when $\epsilon < 0$, and  $F \geq 0.947$ when $\epsilon > 0$. An equivalent investigation with other STA processes was reported in Ref.~\cite{rusch}. Our results show that the robustness of the present scheme to a systematic error in $\Omega_{0}$ is similar to those found with the transitionless CD and inverse engineering schemes. For comparison, here we exhibit the fidelity obtained when a square $\pi$ pulse is used, when the excitation frequency changes from the on-resonance Rabi frequency $\Omega_{R}$ to $\Omega_{R}(1+\epsilon)$, keeping $\Delta =0$, which is given by $F = \sin^2[(1+\epsilon)\pi/2]$. It is important to mention that in experimental realizations, the square $\pi$ pulse is subject to substantial errors in the Rabi frequency. This drawback becomes even more prominent in short time processes, in which high intensity fields must be instantaneously switched on and off at the beginning and the end of the protocol \cite{allen}. In spite of that, the time-rescaling method, which only requires a continuous variation of the fields, is shown to be more robust, principally when $\epsilon >0$. In Fig.~4(b) we see that the time-rescaling method is even more robust against systematic errors in the detuning. In this case, errors of up to 20\% in $\beta^2$ result in $F \geq 0.986$. We are unaware of any similar investigation using other STA methods.    

\subsection*{Thermodynamic cost of the time-rescaled dynamics}
\vspace{0.4 cm}

Unveiling the work cost of a given STA method is also a very important issue \cite{guery,campbell}. Here we compare the properties of the work realized on the system when the reference and TR protocols are applied. For this purpose, we analyze the work cost necessary to cause both types of dynamics using the framework of the ``two-point measurement'' (TPM) protocol \cite{talkner,campisi}. This scheme is useful to describe the energetics of the dynamics of a closed quantum system which may be driven out of equilibrium \cite{batalhao,khan}. Before moving on to this discussion, we stress that the following results are general in the sense that they apply beyond the population inversion scenario to any quantum control dynamics. Let us suppose that the system is initially in thermal equilibrium with a heat bath at a temperature $T$, such that the initial state is a Gibbs thermal state. If we let $\hat{H}_{i}$, $E^{i}_{n}$ and $\ket{n}$ denote the initial Hamiltonian and the respective eigenvalues and eigenkets, the initial state is written as 
\begin{equation}
\label{5}
\hat{\rho}(0) = \sum_{n} \frac{e^{- \beta E^{i}_{n}}}{Z} \ket{n}\bra{n},
\end{equation}
where $Z=$ tr$(e^{-\beta \hat{H}_{i}})$ is the partition function and $\beta = 1/k_{B}T$ the inverse temperature, with $k_{B}$ being the Boltzmann's constant. The first step to capture some information about the work realized on the system is to make a measurement of its energy at $t=0$. In this case, we obtain an outcome $E^{i}_{n}$ with probability $P^{i}_{n} = e^{- \beta E^{i}_{n}}/Z$. 

After this measurement, we immediately disconnect the system from the bath and apply a given reference evolution protocol, $\hat{U}(t_{f})$. At the end of the process, at $t=t_{f}$, we make a second measurement of the energy of the system. At this moment the Hamiltonian is $\hat{H}_{f}$, with $E^{f}_{m}$ and $\ket{m}$ being the respective eigenvalues and eigenkets. The probability to find an outcome $E^{f}_{m}$ in this new measurement is $P^{f}_{n \rightarrow m} = \braket{m|\hat{U}(t_{f})|n}$. Since the system is closed during its evolution, we attribute all energy variation to the work performed in the process from $\ket{n}$ to $\ket{m}$, i.e., $W = E^{f}_{m} - E^{i}_{n}$. Because $E^{f}_{m}$ and $E^{i}_{n}$ may vary in each run of the protocol due to both thermal and quantum mechanical influence, we have that $W$ is a fluctuating quantity. Taking all these elements into account, the work probability distribution in this TPM scheme is \cite{talkner,campisi}
\begin{equation}
\label{13}
P(W) = \sum_{n,m} P^{i}_{n} P^{f}_{n \rightarrow m} \delta[W - (E^{f}_{m} - E^{i}_{n})],
\end{equation}
where $\delta (x)$ is the Dirac delta function.

In many cases of interest, it is cumbersome to work with $P(W)$ due to the large number of possible transitions $\ket{n} \rightarrow \ket{m}$, and hence energy differences $E^{f}_{m} - E^{i}_{n}$. This problem usually becomes simpler if we access the characteristic function given by the Fourier transform of $P(W)$,   
\begin{equation}
\label{14}
\chi(r) = \langle e^{i r W}  \rangle = \int_{-\infty}^{\infty} P(W)e^{i r W} dW.
\end{equation}
By substitution of Eq.~(\ref{13}) into Eq.~(\ref{14}), after some mathematical manipulations, we obtain   
\begin{equation}
\label{15}
\chi(r) = \text{tr} \{\hat{U}^{\dagger}(t_{f}) e^{i r \hat{H}_{f}} \hat{U}(t_{f}) e^{-i r \hat{H}_{i}} \hat{\rho}(0)\}.
\end{equation}
From the definition of $\chi(r)$ in Eq.~(\ref{14}), we can rewrite it in terms of the statistical moments of $W$,
\begin{equation}
\label{16}
\chi(r) = 1 + ir \langle W \rangle - \frac{r^{2}}{2} \langle W^{2} \rangle - i \frac{r^{3}}{3!} \langle W^{3} \rangle + \cdot  \cdot \cdot.
\end{equation}
Now, if we use the expansions $e^{i r \hat{H}_{f}} = 1 + ir \hat{H}_{f} - r^{2}\hat{H}^{2}_{f}/2 + \mathcal{O}(r^{3})$ and $e^{-i r \hat{H}_{i}} = 1 - ir \hat{H}_{i} - r^{2}\hat{H}^{2}_{i}/2 + \mathcal{O}(r^{3})$ in Eq.~(\ref{15}), and compare the result with Eq.~(\ref{16}), we can make the following identifications:
\begin{equation}
\label{17}
\langle W \rangle \equiv \langle \hat{H}_{f}\rangle_{t_{f}} - \langle \hat{H}_{i} \rangle_{0},
\end{equation}

\begin{equation}
\label{18}
\langle W^{2} \rangle  \equiv \langle \hat{H}^{2}_{f}\rangle_{t_{f}} + \langle \hat{H}^{2}_{i} \rangle_{0} - 2\ \text{tr} \{\hat{U}^{\dagger}(t_{f})  \hat{H}_{f} \hat{U}(t_{f})  \hat{H}_{i} \hat{\rho}(0)\},
\end{equation} 
where we have defined the average of a given operator $\hat{A}$ at a time $t$ as $\langle \hat{A} \rangle_{t} = \text{tr} \{\hat{U}^{\dagger}(t)  \hat{A} \hat{U}(t)  \hat{\rho}(0)\}$. Observe from Eq.~(\ref{17}) that $\langle W \rangle $ is the difference between the average energy at $t= t_{f}$
and the average energy at $t=0$, which is not surprising. Still, Eqs.~(\ref{17}) and~(\ref{18}) allow us to calculate the work variance $\langle (\Delta W)^{2} \rangle = \langle W^{2} \rangle-\langle W \rangle^{2}$ as
\begin{align}
\label{19}
\langle (\Delta W)^{2} \rangle  &=  \langle \hat{H}^{2}_{f}\rangle_{t_{f}} + \langle \hat{H}^{2}_{i} \rangle_{0} - 2\ \text{tr} \{\hat{U}^{\dagger} \hat{H}_{f} \hat{U}  \hat{H}_{i} \hat{\rho}(0)\} \nonumber \\
&-  \langle \hat{H}_{f}\rangle^{2}_{t_{f}} - \langle \hat{H}_{i} \rangle^{2}_{0} + 2\ \langle \hat{H}_{f}\rangle_{t_{f}} \langle \hat{H}_{i} \rangle_{0},
\end{align}
where, for simplicity, the time dependence of the evolution operator has been omitted. Here, we can also define the work fluctuation simply as $\Delta W = \sqrt{\langle (\Delta W)^{2} \rangle}$.

Having defined the quantities that characterize the work distribution $P(W)$, we shall compare the results of an arbitrary reference protocol (adiabatic or not) with the corresponding TR (STA or not) process. We observe that the expressions for $\langle W \rangle$ and $\Delta W$ in Eqs.~(\ref{17}) and~(\ref{19}) depend only on the initial and final Hamiltonians, $\hat{H}_{i}$ and $\hat{H}_{f}$, and the evolution operator $\hat{U}$.
Since $\hat{H}_{i}$ and $\hat{H}_{f}$ are the same in the reference and TR dynamics, and the action of $\hat{U}$ is equivalent to the TR evolution operator $\hat{\mathcal{U}}$ of the fast dynamics, we find that \begin{equation}
\label{mean}
\langle W \rangle_{\text{tr}} = \langle W \rangle_{\text{ref}},
\end{equation}     
\begin{equation}
\label{fluctuations}
\Delta W_{\text{tr}} = \Delta W_{\text{ref}}.
\end{equation}
These relations indicate that the mean work done on the system and the work fluctuations are the same for both the reference and TR protocols. 

The results of Eqs.~(\ref{mean}) and~(\ref{fluctuations}) are general, i.e., they are independent of the type of quantum system in which the time-rescaling method is being applied. The average work equality is justified by the fact that the reference and fast dynamics invariably connects the same initial and final states, as expected in any STA method. In turn, the work fluctuation equality is because, despite the correspondence between the initial and finals states, the time-rescaling method also guarantees the same initial and final Hamiltonians. Therefore, the two corresponding energy measurements made in the reference and fast processes have no reason to provide different uncertainties. For the sake of comparison, it was shown in Ref.~\cite{funo} that when CD is used the reference (adiabatic) and fast (STA) work distributions obey $\langle W \rangle_{\text{cd}} = \langle W \rangle_{\text{ref}}$ and $\Delta W_{\text{cd}} > \Delta W_{\text{ref}}$. In that case, the broadening in the work fluctuation is presumably a consequence of the act of switching the auxiliary Hamiltonian on and off in the beginning and the end of the protocol \cite{torrengui}; this concern is absent here.

\subsection*{CONCLUSION}

We studied the problem of speeding up the population inversion
of a two-level quantum system with the recently proposed time-rescaling method \cite{bernardo}, where the  AE adiabatic passage scheme was taken as the reference protocol. The method provides a family of analytical solutions to the time behavior of the Rabi frequency and detuning that are suitable to achieve an arbitrarily fast population inversion. The solutions are characterized as being smooth continuous functions. From the experimental side, it was observed that the TR processes have the advantage of the realization with a single laser. We also examined the robustness of the derived protocols against perturbations in the control parameters and observed that they are as stable as the CD and the inverse engineering methods, with respect to systematic errors in the Rabi frequency. We additionally showed that the stability of the proposed processes is even higher when systematic errors in the detuning are considered. Finally, we investigated the work cost of realizing TR processes and found out that the average and dispersion of the work distributions are unchanged when compared to the reference protocol; a result that holds for any quantum control dynamics. The ideas presented here can be applied to improve the performance of many qubit-based quantum technologies, such as: single-photon emission \cite{miao}, preparation of excitonic qubits with quantum dots \cite{muk}, quantum heat engines \cite{murphy}, and quantum information processing \cite{beterov}. Our work may also provide insights into the control mechanism of photon populations with optical four-wave mixing \cite{ding}. 

\section*{Data availability}

\noindent All data generated or analysed during this study are included in this published article.

\section*{Acknowledgements}

\noindent This work was supported by Coordena{\c c}{\~a}o de Aperfei{\c c}oamento de Pessoal de N{\'i}vel Superior (CAPES, Finance Code 001), and Conselho Nacional de Desenvolvimento Cient{\'i}fico e Tecnol{\'o}gico (CNPq). B.L.B. acknowledges support from (CNPq, Grant No. 303451/2019-0) and PROPESQ/PRPG/UFPB (Project code PIA13177-2020).

\section*{Authors contributions}

\noindent J.S.A. and A.F.S.F. performed the numerical simulations and analyzed the data, B.L.B.  proposed the concept and supervised the work. All authors discussed the results and contributed to the writing and review of the manuscript. 

\section*{Competing interests}

\noindent The authors declare no competing interests.

\section*{Additional information}

\noindent Correspondence and requests for materials should be addressed to B.L.B.

\end{document}